\documentclass[a4paper,11pt]{article}
\pdfoutput=1 

\usepackage{jheppub} 

\usepackage[T1]{fontenc} 

\title{\boldmath Modular functors with infinite dimensional Hilbert spaces}

\author{Takashi Ichikawa}
\affiliation{Department of Mathematics, Faculty of Science and Engineering, 
Saga University, 
\\ 
Saga 840-8502, Japan}

\emailAdd{ichikawn@cc.saga-u.ac.jp}

\abstract{We introduce a notion of generalized modular functors with Hilbert spaces 
of infinite dimension in general, 
and show that a generalized modular functor with data of conformal dimensions determines uniquely wave functions as its flat sections. 
Furthermore, we study an example of generalized modular functors 
derived from the Liouville conformal field theory. 
In this example, wave functions are seen to be gluing conformal blocks 
which are identified with eigenfunctions in the quantum Teichm\"{u}ller theory.}

\begin{document} 
\maketitle
\flushbottom

\section{Introduction}

The notion of modular functors was introduced by Segal \cite{Se} in order to axiomatize  conformal field theory corresponding to vector bundles with projectively flat connection. 
An example of this theory with vector bundles of finite rank was derived from 
affine Lie algebras (cf. Tsuchiya-Ueno-Yamada \cite{TsUY}), 
and another was derived from quantum groups 
(cf. Reshetikhin-Turaev \cite{RT1, RT2} and Turaev \cite{Tu}) 
which were shown to be equivalent by Finkelberg \cite{F} and Andersen-Ueno \cite{AU3}. 

Furthermore, the Liouville conformal field theory (LCFT) and 
the quantum Teichm\"{u}ller theory (QTT) were studied by 
Verlinde \cite{V}, Teschner \cite{T1, T2, T3, T4, T5, T6}, Teschner-Vartanov \cite{TV},  Chekhov-Fock \cite{CF}, Kashaev \cite{K1, K2, K3} and others 
as examples of a generalization of modular functors which associates 
infinite dimensional Hilbert spaces with labeled marked Riemann surfaces. 
Especially, Teschner \cite{T1, T2, T3, T4, T5, T6} and Teschner-Vartanov \cite{TV} 
showed the equivalence of the generalized modular functors from the LCFT and the QTT 
which is related to the AGT correspondence \cite{AGT, AGGTV}. 

In this paper, based on the above researches, we would like to 
\begin{itemize} 

\item 
generalize the notion of modular functors with Hilbert space of infinite dimensional, 
and study Teschner's claim on the Riemann-Hilbert correspondence by showing that 
a generalized modular functor with data of conformal dimensions determines uniquely 
wave functions as its flat sections, 

\item 
show that the LCFT gives a generalized modular functor, 
and hence solve the ``modular functor conjecture'' claimed by Teschner \cite{T3, T5, T6}, 

\item 
study the equivalence between the LCFT and the QTT by using that 
wave functions are given as normalized gluing conformal blocks in the LCFT, 
and as eigenfunctions in the QTT. 

\end{itemize} 
Our main tool to obtain these results is a natural description of the deformation 
from unions of Riemann spheres to Riemann surfaces of positive genus \cite{I2}. 
This tool was applied in \cite{I3} to calculating the monodromy in 
Tsuchiya-Ueno-Yamada's conformal field theory \cite{TsUY}, 
and will be useful in the study of string theory. 

The organization of this paper is as follows. 
In Sections 2, we review the theory of Teichm\"{u}ller groupoids 
\cite{BK1, BK2, FuG, G, HLS, I2, I3, MS, NS} which gives a basis 
of computing exactly the monodromy representations for modular functors. 
In Section 3, 
we introduce the notion of generalized modular functors whose main difference 
from ordinary modular functors is that the factorization isomorphism 
is given by direct integrals of Hilbert spaces.  

In Sections 4, first, 
for a generalized modular functor with data of conformal dimensions 
(i.e., logarithms of Dehn twist eigenvalues), 
we construct a system of Hilbert bundles which satisfies the factorization rule 
defined on the compactified moduli spaces of marked Riemann surfaces. 
Note that a similar construction for original modular functors was given in \cite{BK1, BK2} 
using results of Deligne \cite{D}, 
however Deligne's theory cannot apply in the infinite dimensional case 
since it seems difficult to define the holonomy of connections on general Hilbert bundles. 
Then we construct a canonical local system associated with this modular functor, 
and extend it as a Hilbert bundle on the boundary of the moduli space.    
Second, we show that this Hilbert bundle determines wave functions 
using the notion of tangential base points over ${\mathbb Z}$ at infinity given by 
the arithmetic geometry of Teichm\"{u}ller groupoids \cite{I2, I3}. 

In Section 5, 
we construct a generalized modular functor studied 
by Teschner \cite{T1, T2, T3, T4, T5} from the LCFT 
in which the wave functions are given as normalized gluing conformal blocks. 
Our key observation is that gluing conformal blocks are convergent and analytically continued 
to the moduli space of pointed Riemann surfaces. 
Then by the gluing theory of algebraic curves \cite{I2, I3}, 
the Hilbert spaces of normalized gluing conformal blocks make a generalized modular functor. 
Therefore, we can give a rigorous proof that 
the associated projective representation of basic moves 
(fusing moves, simple moves and half-Dehn twists) in the Teichm\"{u}ller groupoid 
satisfies the Moore-Seiberg equation \cite{MS}.  

In Section 6, 
we review results of Teschner \cite{T3, T4, T6} and Teschner-Vartanov \cite{TV} 
on a conjecture of Verlinde \cite{V} about the equivalence of 
the modular functors derived from the LCFT and the QTT, 
and consider the coincidence of the associated wave functions.

\section{Teichm\"{u}ller groupoid}

\subsection{Degenerate curve} 

We review the correspondence between certain graphs and 
degenerate pointed (algebraic) curves, 
where a (pointed) curve is called {\it degenerate} if it is a stable (pointed) curve and 
the normalization of its irreducible components are all projective (pointed) lines. 
A {\it graph} $\Delta = (V, E, T)$ means a collection  
of 3 finite sets $V$ of vertices, $E$ of edges, $T$ of tails 
and 2 boundary maps 
$$
b : T \rightarrow V, 
\ \ b : E \longrightarrow \left( V \cup \{ \mbox{unordered pairs of elements of $V$} \} \right) 
$$
such that the geometric realization of $\Delta$ is connected. 
A graph $\Delta$ is called {\it stable} 
if its each vertex has degree $\geq 3$, 
i.e. has at least $3$ branches. 
Then for a degenerate pointed curve, 
its dual graph $\Delta = (V, E, T)$ by the correspondence: 
$$
\begin{array}{lcl}
V & \longleftrightarrow & 
\{ \mbox{irreducible components of the curve} \}, 
\\
E & \longleftrightarrow & 
\{ \mbox{singular points on the curve} \}, 
\\
T & \longleftrightarrow & 
\{ \mbox{marked points on the curve} \} 
\end{array}
$$
such that an edge (resp. a tail) of $\Delta$ has a vertex as its boundary 
if the corresponding singular (resp. marked) point belongs 
to the corresponding component. 
Denote by $|X|$ the number of elements of a finite set $X$. 
Under fixing a bijection 
$\nu : T \stackrel{\sim}{\rightarrow} \{ 1, ... , |T| \}$, 
which we call a numbering of $T$, 
a stable graph $\Delta = (V, E, T)$ becomes the dual graph 
of a degenerate $|T|$-pointed curve of genus 
${\rm rank}_{\mathbb Z} H_{1}(\Delta, {\mathbb Z})$ 
and that each tail $h \in T$ corresponds to the $\nu(h)$th marked point. 
In particular, a stable graph without tail is the dual graph of 
a degenerate (non-pointed) curve by this correspondence. 
If $\Delta$ is trivalent, i.e. any vertex of $\Delta$ has just $3$ branches, 
then a degenerate $|T|$-pointed curve with dual graph $\Delta$ 
is maximally degenerate. 

\subsection{Generalized Tate curve} 

Let $\Delta = (V, E)$ be a stable graph without tail, 
and under an orientation of $\Delta$, 
i.e., an orientation of each $e \in E$, 
denote by $v_{h}$ the terminal vertex of $h \in \pm E$ 
(resp. the boundary vertex $b(h)$ of $h \in T)$. 
Take a subset ${\cal E}$ of $\pm E = \{ e, -e \ | \ e \in E \}$ 
whose complement ${\cal E}_{\infty}$ satisfies the condition that 
$$
{\cal E}_{\infty} 
\cap 
\{ -h \ | \ h \in {\cal E}_{\infty} \} 
\ = \ 
\emptyset, 
$$ 
and that $v_{h} \neq v_{h'}$ for any distinct $h, h' \in {\cal E}_{\infty}$. 
We attach variables $\alpha_{h}$ for $h \in {\cal E}$ 
and $q_{e} = q_{-e}$ for $e \in E$. 
Let $A_{0}$ be the ${\mathbb Z}$-algebra generated by 
$\alpha_{h}$ $(h \in {\cal E})$, 
$1/(\alpha_{e} - \alpha_{-e})$ $(e, -e \in {\cal E})$ 
and $1/(\alpha_{h} - \alpha_{h'})$ 
$(h, h' \in {\cal E}$ with $h \neq h'$ and $v_{h} = v_{h'})$, 
and let 
$$ 
A \ = \ A_{0} [[q_{e} \ (e \in E)]], \ \ 
B \ = \ A \left[ \prod_{e \in E} q_{e}^{-1} \right]. 
$$ 
According to \cite[Section 2]{I1}, 
we construct the universal Schottky group $\Gamma$ 
associated with oriented $\Delta$ and ${\cal E}$ as follows. 
For $h \in \pm E$, 
let $\phi_{h}$ be the element of $PGL_{2}(B) = GL_{2}(B)/B^{\times}$ given by 
$$
\phi_{h} \ = \ 
\frac{1}{\alpha_{h} - \alpha_{-h}} 
\left( \begin{array}{cc} 
\alpha_{h} - \alpha_{-h} q_{h} & - \alpha_{h} \alpha_{-h} (1 - q_{h}) 
\\ 1 - q_{h} & -\alpha_{-h} + \alpha_{h} q_{h} 
\end{array} \right) 
\ {\rm mod}(B^{\times}), 
$$
where $\alpha_{h}$ (resp. $\alpha_{-h})$ means $\infty$ 
if $h$ (resp. $-h)$ belongs to ${\cal E}_{\infty}$. 
Then 
$$
\frac{\phi_{h}(z) - \alpha_{h}}{z - \alpha_{h}} 
\ = \ 
q_{h} \frac{\phi_{h}(z) - \alpha_{-h}}{z - \alpha_{-h}} 
\ \ (z \in {\mathbb P}^{1}), 
$$
where $PGL_{2}$ acts on ${\mathbb P}^{1}$ by linear fractional transformation. 
 
For any reduced path $\rho = h(1) \cdot h(2) \cdots h(l)$ 
which is the product of oriented edges $h(1), ... ,h(l)$ 
such that $h(i) \neq -h(i+1)$ and $v_{h(i)} = v_{-h(i+1)}$, 
one can associate an element $\rho^{*}$ of $PGL_{2}(B)$ 
having reduced expression 
$\phi_{h(l)} \phi_{h(l-1)} \cdots \phi_{h(1)}$. 
Fix a base point $v_{b}$ of $V$, 
and consider the fundamental group 
$\pi_{1} (\Delta, v_{b})$ which is a free group 
of rank $g = {\rm rank}_{\mathbb Z} H_{1}(\Delta, {\mathbb Z})$. 
Then the correspondence $\rho \mapsto \rho^{*}$ 
gives an injective anti-homomorphism 
$\pi_{1} (\Delta, v_{b}) \rightarrow PGL_{2}(B)$ 
whose image is denoted by $\Gamma$. 
It is shown in \cite[Section 3]{I1} 
(and had been shown in \cite[Section 2]{IhN} when $\Delta$ is trivalent and has no loop) 
that for any stable graph $\Delta = (V, E)$ without tail, 
there exists a stable curve $C_{\Delta}$ of genus $g$ over $A$ 
which satisfies the following: 

\begin{itemize}

\item 
The closed fiber $C_{\Delta} \otimes_{A} A_{0}$ of $C_{\Delta}$ 
given by putting $q_{e} = 0$ $(e \in E)$ 
is the degenerate curve over $A_{0}$ with dual graph $\Delta$ which is 
obtained from $P_{v} := {\mathbb P}^{1}_{A_{0}}$ $(v \in V)$ 
by identifying $\alpha_{e} \in P_{v_{e}}$ and $\alpha_{-e} \in P_{v_{-e}}$ ($e \in E$), 
where $\alpha_{h} = \infty$ if $h \in {\cal E}_{\infty}$. 

\item 
$C_{\Delta}$ gives a universal deformation of $C_{\Delta} \otimes_{A} A_{0}$. 

\item 
$C_{\Delta} \otimes_{A} B$ is smooth over $B$ 
and is Mumford uniformized (cf. \cite{Mu}) by $\Gamma$. 

\item 
Let $\alpha_{h}$ $(h \in {\cal E})$ be complex numbers 
such that $\alpha_{e} \neq \alpha_{-e}$ 
and that $\alpha_{h} \neq \alpha_{h'}$ if $h \neq h'$ and $v_{h} = v_{h'}$. 
Then for sufficiently small complex numbers $q_{e} \neq 0$ $(e \in E)$, 
$C_{\Delta}$ becomes a Riemann surface which is Schottky uniformized (cf. \cite{S}) 
by $\Gamma$. 

\end{itemize}

We apply the above result to construct a uniformized deformation of 
a degenerate pointed curve which had been done by Ihara and Nakamura \cite[Section 2]{IhN}  when the degenerate pointed curve is maximally degenerate and 
consists of smooth pointed projective lines. 
Let $\Delta = (V, E, T)$ be a stable graph with numbering $\nu$ of $T$. 
We define its extension $\tilde{\Delta} = ( \tilde{V}, \tilde{E} )$ 
as a stable graph without tail by adding a vertex with a loop to the end 
distinct from $v_{h}$ for each tail $h \in T$. 
Then from the uniformized curve associated with $\tilde{\Delta}$, 
by substituting $0$ for the deformation parameters which correspond to 
$e \in \tilde{E} - E$ and by replacing the singular projective lines 
which correspond to $v \in \tilde{V} - V$ with marked points, 
one has the required universal deformation. 

\subsection{Moduli space of curves} 

We review fundamental facts on the moduli space of pointed curves and 
its compactification \cite{DM, KnM, Kn}. 
Let $g$, $n$ be non-negative integers such that $n, 2g - 2 + n > 0$, 
and ${\cal M}_{g, n}$ (resp. ${\cal M}_{g, \vec{n}}$) denote the moduli stacks 
over ${\mathbb Z}$ of proper smooth curves of genus $g$ with $n$ marked points 
(resp. $n$ non-zero tangent vectors). 
Then ${\cal M}_{g, \vec{n}}$ becomes naturally a principal $({\mathbb G}_{m})^{n}$-bundle 
on ${\cal M}_{g, n}$. 
Furthermore, let $\overline{\cal M}_{g, n}$ denote the Deligne-Mumford-Knudsen 
compactification of ${\cal M}_{g, n}$ which is defined as the moduli stack 
over ${\mathbb Z}$ of stable curves of genus $g$ with $n$ marked points, 
and $\overline{\cal M}_{g, \vec{n}}$ denote the $({\mathbb A}^{1})^{n}$-bundle 
on $\overline{\cal M}_{g, n}$ containing ${\cal M}_{g, \vec{n}}$ naturally. 
For these moduli stacks ${\cal M}_{*,*}$ and $\overline{\cal M}_{*,*}$, 
${\cal M}_{*,*}^{\rm an}$ and $\overline{\cal M}_{*,*}^{\rm an}$ denote 
the associated complex orbifolds. 
A {\it point at infinity} on ${\cal M}_{g, n}$ (resp. ${\cal M}_{g, \vec{n}}$) is a point on 
$\overline{\cal M}_{g, n}$ (resp. $\overline{\cal M}_{g, \vec{n}}$) which corresponds to 
a maximally degenerate $n$-pointed curve, 
and a {\it tangential point over ${\mathbb Z}$ at infinity} is 
a point at infinity with tangential structure over ${\mathbb Z}$. 

We describe the boundary of ${\cal M}_{g, n}$. 
Denote by ${\cal D}_{0}$ the divisor of $\overline{\cal M}_{g, n}$ consisting of 
stable pointed curves which are obtained from stable curves of $g-1$ 
and $n+2$ marked points by identifying the $(n+1)$th and $(n+2)$th points. 
Let $1 \leq i \leq [g/2]$ be an integer, 
and $S$ be a subset of $\{ 1,..., n \}$ such that $2i - 2 + |S|$, $2(g-i) - 2 + n - |S|$ 
are positive. 
Then the divisor ${\cal D}_{i,S}$ of $\overline{\cal M}_{g, n}$ consists of stable pointed curves  which are obtained from pairs of stable curves of genus $i$ (resp. $g-i$) with marked points 
indexed by $S \cup \{ n+1 \}$ (resp. $(\{ 1,..., n \} - S) \cup \{ n+2 \}$) 
by identifying their points indexed by $n+1$ and $n+2$. 
Then $\overline{\cal M}_{g, n} - {\cal M}_{g, n}$ consists of normal crossing divisors 
${\cal D}_{0}, {\cal D}_{i,S}$, 
and hence $\overline{\cal M}_{g, \vec{n}} - {\cal M}_{g, \vec{n}}$ consists of 
the pullbacks of ${\cal D}_{0}, {\cal D}_{i,S}$ by the natural projection 
$\overline{\cal M}_{g, \vec{n}} \rightarrow \overline{\cal M}_{g, n}$ 
which we denote by the same notation. 

\subsection{Local coordinates on the moduli space} 

To describe local coordinates on the moduli stack of stable pointed curves 
using the universal deformation, 
we will rigidify a coordinate on each projective line appearing 
as an irreducible component of the base degenerate curve. 
In the maximally degenerate case, this process is considered in \cite{IhN} 
using the notion of ``tangential structure''. 
A rigidification of an oriented stable graph $\Delta = (V, E, T)$ with numbering
$\nu$ of $T$ means a collection $\tau = \left( \tau_{v} \right)_{v \in V}$ 
of injective maps
$$
\tau_{v} : \{ 0, 1, \infty \} \rightarrow 
\left\{ h \in \pm E \cup T \ | \ v_{h} = v \right\} 
$$
such that $\tau_{v}(a) \neq - \tau_{v'}(a)$ for any $a \in \{ 0, 1, \infty \}$ 
and distinct elements $v, v' \in V$ with $\tau_{v}(a), \tau_{v'}(a) \in \pm E$. 
One can see that any stable graph has a rigidification by the induction
on the number of edges and tails. 
Denote by $A_{(\Delta, \tau)}$ the ring of formal power series ring of $q_{e}$ $(e \in E)$  over the ${\mathbb Z}$-algebra which is generated by 
$\alpha_{h}$ ($h \in {\cal E}$), $1/(\alpha_{e} - \alpha_{-e})$ ($e, -e \in {\cal E} - T$) 
and $1/(\alpha_{h} - \alpha_{h'})$ ($h, h' \in {\cal E}$ with $h \neq h'$ and $v_{h} = v_{h'}$), 
where  
$$ 
{\cal E} = \pm E \cup T - \left\{ \tau_{v}(\infty) \ | \ v \in V \right\}, 
$$
$\alpha_{h} = a$ for $h = \tau_{v}(a)$ $(a \in \{ 0, 1 \})$ and $\alpha_{h}$ are variables. 
Then as is stated above, 
there exists a stable $|T|$-pointed curve $C_{(\Delta, \tau)}$ over 
$A_{(\Delta, \tau)}$ which is obtained as the quotient by $\pi_{1}(\Delta)$ 
of the glued scheme of pointed projective lines associated with 
the universal cover of $\Delta$. 
Therefore, $C_{(\Delta, \tau)}$ gives a universal deformation by $q_{e}$ $(e \in E)$ of 
the degenerate $|T|$-pointed curve with dual graph $\Delta$. 

If one takes another rigidification, 
then $q_{e}$ $(e \in E)$ may give different deformation parameters of 
the degenerate pointed curve, 
and these parameters associated with distinct rigidifications can be compared 
in \cite[Section 2]{I2} 
by using the above theory of Schottky-Mumford uniformization. 

Let $\tau$ be a rigidification of an oriented stable graph 
$\Delta = (V, E, T)$ with numbering of $T$, 
and put
$$
{\cal E}_{\tau} = \pm E \cup T - \bigcup_{v \in V} {\rm Im}(\tau_{v}). 
$$
Then $\alpha_{h}$ $(h \in {\cal E}_{\tau})$ and $q_{e}$ $(e \in E)$ give effective parameters 
of the moduli and the deformation of degenerate $|T|$-pointed curves 
with dual graph $\Delta$ respectively. 
Therefore, if we put $g = {\rm rank}_{\mathbb Z} H_{1} (\Delta, {\mathbb Z})$ 
and $n = |T|$, 
then
$$
\left( \alpha_{h} \ (h \in {\cal E}_{\tau}), q_{e} \ (e \in E) \right) 
$$
gives a system of formal coordinates on an \'{e}tale neighborhood of $Z_{\Delta}$, 
where $Z_{\Delta}$ denotes the substack of $\overline{\cal M}_{g, n}$ 
classifying  degenerate $n$-pointed curves with dual graph $\Delta$. 
Furthermore, by the above result, 
this system gives local coordinates on an \'{e}tale neighborhood of 
the complex orbifold $Z_{\Delta}^{\rm an}$ associated with $Z_{\Delta}$. 

\subsection{Teichm\"{u}ller groupoid} 

The {\it Teichm\"{u}ller groupoid} for ${\cal M}_{g, \vec{n}}$ is defined as 
the fundamental groupoid for ${\cal M}_{g, \vec{n}}^{\rm an}$ 
with tangential base points over ${\mathbb Z}$ at infinity. 
Its fundamental paths called {\it basic moves} are half-Dehn twists, 
fusing moves and simple moves defined as follows. 

Let $\Delta = (V, E, T)$ be a stable graph as above, 
and assume that $\Delta$ is trivalent. 
Then for any rigidification $\tau$ of $\Delta$, 
$\pm E \cup T = \bigcup_{v \in V} {\rm Im}(\tau_{v})$, 
and hence $A_{(\Delta, \tau)}$ is the formal power series ring over ${\mathbb Z}$ 
of $3g + n - 3$ variables $q_{e}$ $(e \in E)$. 
First, the {\it half-Dehn twist} $\delta_{e}^{1/2}$ associated with $e$ is defined 
as the deformation of the pointed Riemann surface corresponding to $C_{\Delta}$ 
by $q_{e} \mapsto - q_{e}$. 
Second, a {\it fusing move (or associative move, A-move)} is defined to be 
different degeneration processes of a $4$-hold Riemann sphere. 
A fusing move $\varphi(e, e')$ changes $(\Delta, e)$ to another trivalent graph 
$(\Delta', e')$ such that $\Delta$, $\Delta'$ become the same graph, 
which we denote by $\Delta''$, if $e, e'$ shrink to a point. 
In \cite[Section 2]{I2}, $\varphi(e, e')$ is constructed using $C_{\Delta''}$. 
Finally,  {\it simple move (or S-move)} is defined to be different degeneration processes 
of a $1$-hold complex torus. 

Then as the {\it completeness theorem} called by Moore-Seiberg \cite{MS}, 
the following Theorem 2.1 was conjectured by Grothendieck \cite{G} and shown by 
Bakalov-Kirillov, Funar-Gelca, Hatcher-Lochak-Schneps \cite{BK1, BK2, FuG, HLS}, 
especially by Nakamura-Schneps \cite[Sections 7 and 8]{NS} 
using the notion of quilt-decompositions of Riemann surfaces. 
\vspace{2ex}

\noindent
{\bf Theorem 2.1} (cf. \cite{BK1, BK2, FuG, G, HLS, MS} and 
\cite[Sections 7 and 8]{NS}).    
\begin{it}
The Teichm\"{u}ller groupoid is generated by half-Dehn twists, 
fusing moves and  simple moves with relations induced from 
${\cal M}_{0, \vec{4}}$, ${\cal M}_{0, \vec{5}}$, ${\cal M}_{1, \vec{1}}$ 
and ${\cal M}_{1, \vec{2}}$. 
\end{it}

\section{Modular functor} 

\subsection{Marked surface} 

Let $\Sigma$ be a closed oriented (real) surface $\Sigma$ of genus $g$ 
which is not necessarily connected, 
and denote by 
$$
\left( \cdot, \cdot \right) : 
H_{1} \left( \Sigma, {\mathbb Z} \right) \times H_{1} \left( \Sigma, {\mathbb Z} \right) 
\rightarrow {\mathbb Z} 
$$
the non-degenerate skew-symmetric intersection pairing. 
If $\Sigma$ is connected, 
then a Lagrangian subspace $L \subset H_{1} \left( \Sigma, {\mathbb Z} \right)$ means  a subspace which is maximally isotropic with respect to the intersection pairing. 
If $\Sigma$ is not connected and 
$H_{1} \left( \Sigma, {\mathbb Z} \right) = 
\oplus_{i} H_{1} \left( \Sigma, {\mathbb Z} \right)$, 
where $\Sigma_{i}$ are the connected components of $\Sigma$, 
then a Lagrangian subspace is a subspace of the form 
$L = \oplus_{i} L_{i}$, 
where each $L_{i} \subset H_{1} \left( \Sigma, {\mathbb Z} \right)$ is Lagrangian. 
For any real vector space $V$, 
put ${\mathbb P}(V) = (V - \{ 0 \})/{\mathbb R}_{+}$, 
where
$$
{\mathbb R}_{+} = \{ r \in {\mathbb R} \ | \ r > 0 \}. 
$$  

A {\it marked surface} ${\bf \Sigma} = (\Sigma, P, V, L)$ is 
an oriented closed smooth surface $\Sigma$ with a finite subset $P$ of points on $\Sigma$, 
an element $V$ of $\sqcup_{p \in P} {\mathbb P} \left( T_{p} \Sigma \right)$  
consisting of projective tangent vectors at $p \in P$ 
and a Lagrangian subspace $L \subset H_{1} \left( \Sigma, {\mathbb Z} \right)$. 
A {\it morphism} ${\bf f} = (f, s): {\bf \Sigma}_{1} \rightarrow {\bf \Sigma}_{2}$ 
of marked surfaces ${\bf \Sigma}_{i} = (\Sigma_{i}, P_{i}, V_{i}, L_{i})$ 
is an isotopy class of orientation preserving diffeomorphisms 
$f : \Sigma_{1} \rightarrow \Sigma_{2}$ which maps $(P_{1}, V_{1})$ to $(P_{2}, V_{2})$ 
together with an integer $s$. 
Let ${\bf f}_{1} = (f_{1}, s_{1}) : {\bf \Sigma}_{1} \rightarrow {\bf \Sigma}_{2}$ and 
${\bf f}_{2} = (f_{2}, s_{2}) : {\bf \Sigma}_{2} \rightarrow {\bf \Sigma}_{3}$ 
be morphisms of marked surfaces ${\bf \Sigma}_{i} = (\Sigma_{i}, P_{i}, V_{i}, L_{i})$. 
Then the composition of ${\bf f}_{1}$ and ${\bf f}_{2}$ is 
$$
{\bf f}_{2} {\bf f}_{1} = \left( f_{2} \circ f_{1}, s_{2} + s_{1} - \sigma 
\left( (f_{2} \circ f_{1})_{*} L_{1}, (f_{2})_{*} L_{2}, L_{3} \right) \right), 
$$ 
where $\sigma$ denotes the Wall signature cocycle for triples of 
Lagrangian subspaces of $H_{1} \left( \Sigma, {\mathbb R} \right)$. 
Then one can define the category of marked surfaces. 

The mapping class group $\Gamma({\bf \Sigma})$ of a marked surface 
${\bf \Sigma} = (\Sigma, P, L)$ is the group of automorphisms of ${\bf \Sigma}$. 
One can see that $\Gamma({\bf \Sigma})$ is a central extension of 
the framed mapping class group $\Gamma(\Sigma, P)$ 
of the pointed surface $(\Sigma, P)$: 
$$
0 \rightarrow {\mathbb Z} \rightarrow \Gamma({\bf \Sigma}) \rightarrow 
\Gamma(\Sigma, P) \rightarrow 0 
$$
which is defined by the $2$-cocycle $c$ on $\Gamma({\bf \Sigma})$, 
where 
$$
c(f_{1}, f_{2}) = \sigma \left( (f_{1} \circ f_{2})_{*} L, (f_{1})_{*} L, L \right). 
$$

\subsection{Axiom of generalized modular functors} 

We give the axioms for a $2$-dimensional {\it generalized} modular functor extending 
the series of papers by Andersen-Ueno \cite{AU1, AU2, AU3}. 

A {\it label set} $\Lambda$ is a measure space which has an involution 
$\lambda \mapsto \hat{\lambda}$. 
The category of {\it $\Lambda$-labeled marked surfaces} consists of 
marked surfaces with an element of $\Lambda$ attached to each of the marked points, 
and morphisms of labeled marked surfaces are required to preserve the labelings. 
We denote the labeled marked surface by $({\bf \Sigma}, \vec{\lambda})$, 
where $\vec{\lambda}$ denotes the labelling.
A {\it generalized modular functor} for the label set $\Lambda$ is a functor $V$ 
from the category of labeled marked surfaces to the category of 
(complex) Hilbert spaces which satisfies the following axioms. 
\vspace{2ex}

\noindent
{\bf Disjoint union axiom}. 
The operation of disjoint union of labeled marked surfaces corresponds to 
the operation of tensor product, 
i.e. for any pair of labeled marked surfaces there exists an isomorphism
$$
V (({\bf \Sigma}_{1}, \vec{\lambda}_{1}) \sqcup ({\bf \Sigma}_{2}, \vec{\lambda}_{2})) \cong 
V ({\bf \Sigma}_{1}, \vec{\lambda}_{1}) \otimes V ({\bf \Sigma}_{2}, \vec{\lambda}_{2}) 
$$
which is associative.
\vspace{2ex}

\noindent
{\bf Gluing axiom}. 
For a marked surface 
${\bf \Sigma} = \left( \Sigma, \{ p_{1}, p_{2} \} \sqcup P, \{ v_{1}, v_{2} \} \sqcup V, L \right)$, 
let ${\bf \Sigma}_{c} = \left( \Sigma_{c}, P, V, L_{c} \right)$ 
denote the marked surface obtained by gluing $(p_{1}, v_{1}), (p_{2}, v_{2})$ 
via an orientation reversing projective linear isomorphism 
$c : {\mathbb P} \left( T_{p_{1}} \Sigma \right) \rightarrow 
{\mathbb P} \left( T_{p_{2}} \Sigma \right)$ 
such that $c(v_{1}) = v_{2}$. 
Furthermore, 
$L_{c} = q_{*}^{-1} \left( n_{*}(L) \right)$, 
where $q : \Sigma_{c} \rightarrow \Sigma'$ and $n : \Sigma \rightarrow \Sigma'$ are 
the natural continuous maps to the singular surface $\Sigma'$ obtained from $\Sigma$ 
by identifying $p_{1} = p_{2}$. 
Then there exists an isomorphism between $V ({\bf \Sigma}_{c}, \vec{\lambda})$ and 
the direct integral of $V ({\bf \Sigma}, (\vec{\lambda}, \mu, \hat{\mu}))$ 
by the measure $d \mu$ $(\mu \in \Lambda)$, namely 
$$
V ({\bf \Sigma}_{c}, \vec{\lambda}) \cong 
\int_{\Lambda}^{\oplus} d \mu \ V ({\bf \Sigma}, (\vec{\lambda}, \mu, \hat{\mu})). 
$$
Furthermore, this isomorphism, 
which is called the gluing isomorphism or the factorization isomorphism, 
is associative, compatible with gluing of morphisms, 
disjoint unions and it is independent of the choice of the gluing map in the obvious way. 
\vspace{2ex}

\noindent
{\bf Twice punctured sphere axiom}. 
Let ${\bf \Sigma} = \left( S^{2}, \{ p_{1}, p_{2} \}, \{ v_{1}, v_{2} \}, \{ 0 \} \right)$ 
be a $2$-pointed marked sphere. 
Then
$$
\dim_{\mathbb C} V ({\bf \Sigma}, (\lambda, \mu)) = 
\left\{ \begin{array}{ll} 
1 & (\lambda = \hat{\mu}), \\ 0 & (\lambda \neq \hat{\mu}). 
\end{array} \right. 
$$

\section{Equivalence of generalized modular functors} 

\subsection{Conformal dimension} 

Let $\left\{ V ({\bf \Sigma}, \vec{\lambda}) \right\}_{{\bf \Sigma}, \vec{\lambda}}$ 
be a generalized modular functor for the label set $\Lambda$. 
Then by factorizing $({\rm id}_{\Sigma}, 1)$ as 
$({\rm id}_{\Sigma_{0}}, 1) \cup ({\rm id}_{\Sigma - \Sigma_{0}}, 0)$ 
along the boundary of an disc $\Sigma_{0}$ embedded in $\Sigma - P$, 
one can see that  
$$
V ({\rm id}_{\bf \Sigma}, 1) \in {\rm Aut} \left( V({\bf \Sigma}, \vec{\lambda}) \right) 
$$
is given as multiplication by a scalar $\tilde{c} \in {\mathbb C}^{\times}$ 
independent of ${\bf \Sigma}$ and $\vec{\lambda}$. 
Furthermore, 
for the Dehn twist $\delta$ along the equator on a $2$-pointed sphere 
${\bf \Sigma} = \left( S^{2}, \{ p_{1}, p_{2} \} \right)$,  
$$
V(\delta) \in {\rm Aut} \left( V({\bf \Sigma}, (\lambda, \hat{\lambda}) \right) 
$$
is given as multiplication by a scalar $\tilde{r}_{\lambda} \in {\mathbb C}^{\times}$ 
satisfying  $\tilde{r}_{\lambda} = \tilde{r}_{\hat{\lambda}}$.  
Then the action of 
$$
\mbox{$V(\delta) = \tilde{r}_{\mu} \cdot {\rm id.}$ on 
$V ({\bf \Sigma}, (\vec{\lambda}, \mu, \hat{\mu}))$}
$$ 
gives a representation of a Dehn twist as an automorphism 
on the Hilbert space $V ({\bf \Sigma}_{c},  \vec{\lambda})$ 
under the gluing isomorphism. 
A set of {\it conformal dimensions} consists of complex numbers $r_{\lambda}$ 
$(\lambda \in \Lambda)$ such that ${\bf e}(r_{\lambda}) = \tilde{r}_{\lambda}$ 
and $r_{\lambda} = r_{\hat{\lambda}}$, 
where 
$$
{\bf e}(z) := \exp \left( 2 \pi \sqrt{-1} z \right). 
$$

\noindent
{\bf Proposition 4.1.} 
\begin{it} 
Let $\{ r_{\lambda} \}_{\lambda \in \Lambda}$ be a set of conformal dimensions. 
Then for a complex number $z \neq 0$, 
the multiplication by $\exp \left( r_{\mu} \cdot \log(z) \right) |z|^{-r_{\mu}}$ 
on $V(\mbox{\boldmath $\Sigma$}, (\vec{\lambda}, \mu, \hat{\mu}))$ 
$(\lambda \in \Lambda)$ give an automorphism on the Hilbert space 
$V(\mbox{\boldmath $\Sigma$}_{c}, \vec{\lambda})$ 
under the factorization isomorphism. 
\end{it}
\vspace{2ex}

\noindent
{\it Proof.} 
Take a positive integer $m$ such that ${\rm arg}(z)/(2 \pi) \leq m$. 
Then the assertion follows from that 
$\left| \exp \left( r_{\mu} \cdot \log(z) \right) |z|^{-r_{\mu}} \right| 
\leq \left| \tilde{r}_{\mu} \right|^{m}$ 
and that the multiplication by $\tilde{r}_{\mu}^{m}$ 
on $V(\mbox{\boldmath $\Sigma$}, (\vec{\lambda}, \mu, \hat{\mu}))$ 
$(\lambda \in \Lambda)$ give an automorphism on $V(\mbox{\boldmath $\Sigma$}_{c}, \vec{\lambda})$ . 
\ $\square$ 

\subsection{Hilbert bundle} 

From a generalized modular functor, 
we construct projective local system of Hilbert spaces on the moduli spaces of curves  following arguments in Andersen-Borot-Orantin \cite[2.5]{ABO}, 
and extend these bundles to the compactified moduli spaces. 

We consider the trivial local system 
${\cal T}(\Sigma, P) \times V({\bf \Sigma}, \vec{\lambda})$ 
on the Teichm\"{u}ller space ${\cal T}(\Sigma, P)$ 
for $(\Sigma, P)$. 
Let ${\cal L}_{\rm H}$ be the Hodge line bundle, 
and take a complex number (called the {\it central charge}) $c$ 
such that $\tilde{c} = {\bf e}(c/4)$. 
Then using the section $u_{1} \wedge \cdots \wedge u_{g}$ of ${\cal L}_{\rm H}$ 
induced from integral basis $(u_{1},..., u_{g})$ of $L$, 
one can define ${\cal L}_{\rm H}^{-c/2}$ on which $({\rm id}, 1)$ acts by $\tilde{c}^{-1}$. 
Therefore, by \cite[Theorem 2.5]{ABO}, 
the quotient of 
$$
{\cal T}(\Sigma, P) \times V({\bf \Sigma}, \vec{\lambda}) \times {\cal L}_{\rm H}^{-c/2} 
$$ 
by the action of $\Gamma(\Sigma, P)$ can be defined as 
a projective local system of Hilbert spaces on 
${\cal T}(\Sigma, P)/\Gamma(\Sigma, P) \cong {\cal M}_{g, \vec{n}}^{\rm an}$, 
where $g$ and $n$ are the genus and the number of marked points of $\Sigma$ respectively. 
Then we denote by ${\cal V}(g, n, \vec{\lambda})$ 
this local system and the associated Hilbert bundle. 
\vspace{2ex}

\noindent
{\bf Theorem 4.2.} 
\begin{it} 
Let $\{ r_{\lambda} \}_{\lambda \in \Lambda}$ be a set of conformal dimensions. 
Then there exists a system 
$$
\overline{\cal V} = 
\left\{ \overline{\cal V}(g, n, \vec{\lambda}) \right\}_{g, n, \vec{\lambda}}
$$
of Hilbert bundles $\overline{\cal V}(g, n, \vec{\lambda})$ 
on $\overline{\cal M}_{g, \vec{n}}^{\rm an}$ which is an extension of 
${\cal V}(g, n, \vec{\lambda})$ 
and satisfies the following conditions: 
\begin{itemize}

\item[{\rm (1)}] 
the fiber of $\overline{\cal V}(g, n, \vec{\lambda})$ around ${\cal D}_{0}$ 
is associated with the space 
$$
\int_{\Lambda}^{\oplus} d \mu \ {\cal V}(g-1, n+2, (\vec{\lambda}, \mu, \hat{\mu}))
$$
under the factorization isomorphism
$$
{\cal V}(g, n, \vec{\lambda}) \cong \int_{\Lambda}^{\oplus} 
d \mu \ {\cal V}(g-1, n+2, (\vec{\lambda}, \mu, \hat{\mu}))
$$
on a neighborhood of ${\cal D}_{0}$ in ${\cal M}_{g, \vec{n}}^{\rm an}$. 

\item[{\rm (2)}] 
the fiber of $\overline{\cal V}(g, n, \vec{\lambda})$ around ${\cal D}_{i,m}$ 
is associated with the space 
$$
\int_{\Lambda}^{\oplus} d \mu \ {\cal V}(i, S, (\vec{\lambda}', \mu)) 
\otimes {\cal V}(g-i, \{ 1,..., n \}-S, (\vec{\lambda}'', \hat{\mu})), 
$$
where $\vec{\lambda}'$ and $\vec{\lambda}''$ are sublabellings of $\vec{\lambda}$ 
corresponding to the partition of $\{ 1,..., n \}$ to $S$ and $\{ 1,..., n \}-S$ respectively, 
under the factorization isomorphism
$$
{\cal V}(g, n, \vec{\lambda}) \cong \int_{\Lambda}^{\oplus} 
d \mu \ {\cal V}(i, S, (\vec{\lambda}', \mu)) 
\otimes {\cal V}(g-i, \{ 1,..., n \}-S, (\vec{\lambda}'', \hat{\mu}))
$$
on a neighborhood of ${\cal D}_{i, m}$ in ${\cal M}_{g, \vec{n}}^{\rm an}$. 
\end{itemize}
\end{it}

\noindent
{\it Proof.} 
Let $p$ be a point on 
$\partial {\cal M}_{g, n}^{\rm an} = 
\overline{\cal M}_{g, n}^{\rm an} - {\cal M}_{g, n}^{\rm an}$. 
Then there are an integer $k$ with $0 \leq k < 3g - 3 + n$ and local coordinates 
$(z_{1},..., z_{3g-3+n})$ on a neighborhood $U_{p}$ of $p$ 
such that ${\cal M}_{g, n}^{\rm an} \cap U_{p}$ is 
$U'_{p} = \left\{ (z_{i}) \in U_{p} \ \left| \ \prod_{i=1}^{k} z_{i} \neq 0 \right. \right\}$. 
The factorization isomorphism identifies $V(\mbox{\boldmath $\Sigma$}, \vec{\lambda})$ 
with the direct integral $V(\mbox{\boldmath $\Sigma$}, \vec{\lambda})_{p}$ 
by the measures $d \mu_{i}$ $(1 \leq i \leq k)$ of conformal blocks 
$V(\mbox{\boldmath $\Sigma$}', (\vec{\lambda}', \mu, \hat{\mu}))$ 
associated with the stable curve corresponding to $p$. 
By the axioms in 3.2 and Proposition 4.1, 
the multiplication by  
$$
\prod_{i=1}^{k} \exp \left( r_{\mu} \cdot \log(z_{i}) \right) |z_{i}|^{-r_{\mu}}
$$
on each factor $V(\mbox{\boldmath $\Sigma$}', (\vec{\lambda}', \mu, \hat{\mu}))$ 
gives an automorphism on $V(\mbox{\boldmath $\Sigma$}, \vec{\lambda})_{p}$.  
Then we define a Hilbert bundle on $U_{p}$ (uniquely determined up to an isomorphism)  as the trivial bundle with fiber $V(\mbox{\boldmath $\Sigma$}, \vec{\lambda})_{p}$ 
which has the structure of a local system on the universal cover of $U'_{p}$ 
obtained by this automorphism of $V(\mbox{\boldmath $\Sigma$}, \vec{\lambda})_{p}$.  
The monodromy around $p$ of this local system is same to that of  
${\cal V}(g, n, \vec{\lambda})$, 
and the factorization isomorphisms are compatible with degeneration processes. 
Therefore, by gluing ${\cal V}(g, n, \vec{\lambda})$ with these Hilbert bundles, 
we have its extension $\overline{\cal V}(g, n, \vec{\lambda})$ as a Hilbert bundle 
on $\overline{\cal M}_{g, \vec{n}}^{\rm an}$ satisfying the required conditions. 
\ $\square$ 

\subsection{Wave function} 

\noindent
{\bf Theorem 4.3.} 
\begin{it} 
For a set $\{ r_{\lambda} \}_{\lambda \in \Lambda}$ of conformal dimensions, 
denote by $\overline{\cal V}$ the Hilbert bundle on $\overline{\cal M}_{g, \vec{n}}^{\rm an}$ 
constructed in Theorem 4.2. 
Let $p_{\infty}$ be a tangential base point over ${\mathbb Z}$ at infinity 
obtained by taking the parameters $q_{e}$ $(e \in E)$ in 2.4 
as sufficiently small positive numbers. 
Then for a $\Lambda$-valued function $\beta$ on $E$, 
and an element $\alpha$ of the fiber of $\overline{\cal V}$ at $p_{\infty}$,  
there exists uniquely the wave function ${\cal F}$ with initial value $\alpha$ 
as a flat section around $p_{\infty}$ of $\overline{\cal V}$ such that 
${\displaystyle \lim_{q_{e} \downarrow 0} {\cal F} = \alpha}$. 
\end{it} 
\vspace{2ex}

\noindent
{\it Proof.} 
The assertion follows from the construction of $\overline{\cal V}$ in Theorem 4.2. 
\ $\square$

\section{Liouville conformal field theory}  

\subsection{Conformal block} 

Fix a real number $c > 1$ called the {\it Liouville central charge}, 
and define the {\it Virasoro algebra} ${\rm Vir}_{c}$ with generators 
$L_{n}$ $(n \in {\mathbb Z})$ satisfying the relations 
$$
[L_{n}, L_{m}] = (n-m) L_{n+m} + \frac{c}{12} n (n^{2} - 1) \delta_{n+m, 0}, 
$$
where $\delta_{i, j}$ denotes the Kronecker delta. 
Take a real number $Q$ such that $c = 1 + 6 Q^{2}$. 
Then the set 
${\mathbb S} = \left. \left\{ Q/2 + \sqrt{-1} r \ \right| \ r > 0 \right\}$ 
parametrizes irreducible highest weight representations of ${\rm Vir}_{c}$ 
called {\it Verma modules}. 
More precisely, for each $\alpha \in {\mathbb S}$, 
there is an irreducible highest weight representation ${\cal V}_{\alpha}$ of ${\rm Vir}_{c}$ 
with generator $e_{\alpha}$ which is annihilated by $L_{n}$ $(n > 0)$ and 
has the $L_{0}$-eigenvalue $\Delta_{\alpha} = \alpha (Q - \alpha)$. 
Then there exists a unique inner product 
$\langle \cdot, \cdot \rangle_{{\cal V}_{\alpha}}$ on ${\cal V}_{\alpha}$ such that 
$$
\left\langle L_{n}(v), w \right\rangle_{{\cal V}_{\alpha}} = 
\left\langle v, L_{-n}(w) \right\rangle_{{\cal V}_{\alpha}}     
\ (v, w \in {\cal V}_{\alpha}) 
$$
and that $\langle e_{\alpha}, e_{\alpha} \rangle_{{\cal V}_{\alpha}} = 1$. 
Note that for any $v \in {\cal V}_{\alpha}$, $L_{n}(v) = 0$ if $n \gg 0$. 

Under $2g - 2 + n > 0$, 
let $C$ be a Riemann surface of genus $g$ with $n$ marked points $p_{1},..., p_{n}$ 
and local coordinates $z_{i}$ $(1 \leq i \leq n)$ vanishing at $p_{i}$. 
We associate highest weight representations ${\cal V}_{\alpha_{i}}$ of ${\rm Vir}_{c}$ to $p_{i}$, 
and define the action of 
$$
\chi = \left( \sum_{k \in {\mathbb Z}} \chi_{k}^{(i)} z_{i}^{k+1} \partial_{z_{i}} 
\right)_{1 \leq i \leq n} \in \bigoplus_{i=1}^{n} {\mathbb C}((z_{i})) \partial_{z_{i}}
$$ 
on $\bigotimes_{i=1}^{n} {\cal V}_{\alpha_{i}}$ as the following finite sum  
$$
\rho_{\chi} (v_{1} \otimes \cdots \otimes v_{n}) = 
\sum_{i=1}^{n} v_{1} \otimes \cdots \otimes 
\left( \sum_{k \in {\mathbb Z}} \chi_{k}^{(i)} L_{k}(v_{i}) \right) 
\otimes \cdots \otimes v_{n} \ \left( v_{i} \in {\cal V}_{\alpha_{i}} \right). 
$$
Denote by ${\mathfrak D}_{C}$ the Lie algebra of meromorphic differential operators 
on $C$ which are holomorphic except $p_{1},..., p_{n}$. 
Then (invariant) conformal blocks associated with $(C; p_{i}, z_{i})$ are linear maps 
${\cal F}_{C} : \bigotimes_{i=1}^{n} {\cal V}_{\alpha_{i}} \rightarrow {\mathbb C}$ 
satisfying the invariance property: 
$$
{\cal F}_{C}(\rho_{\chi}(v)) = 0 \ \ \left( \chi \in {\mathfrak D}_{C}, \ 
v \in \bigotimes_{i=1}^{n} {\cal V}_{\alpha_{i}} \right),
$$
where $\chi$ is regarded as an element of 
$\bigoplus_{i=1}^{n} {\mathbb C}((z_{i})) \partial_{z_{i}}$ (cf. \cite{T3,T5}).  
If $(g, n) = (0, 3)$, 
then ${\cal F}_{C}$ is uniquely determined by the values 
${\cal F}_{C} \left( e_{\alpha_{1}} \otimes e_{\alpha_{2}} \otimes e_{\alpha_{3}} \right)$. 

Let $C_{k}$ $(k = 1, 2)$ be two Riemann surfaces with $n_{k} + 1$ marked points 
and associated local coordinates. 
Then we denote by $C_{1} \sharp C_{2}$ the $(n_{1} + n_{2})$-pointed Riemann surface 
obtained by gluing $C_{k}$ at their $(n_{k} + 1)$th points with local coordinates $z_{n_{k}+1}$ 
via the gluing parameter $q$, 
namely $z_{n_{1}+1} \cdot z_{n_{2}+1} = q$ on $C_{1} \sharp C_{2}$. 
The gluing of conformal blocks ${\cal F}_{C_{1}}, {\cal F}_{C_{2}}$ by $\beta \in {\mathbb S}$ 
is defined as 
$$
{\cal F}_{C_{1} \sharp C_{2}}^{\beta} (v_{1} \otimes v_{2}) = 
\sum_{l,m} {\cal F}_{C_{1}} (v_{1} \otimes v_{l}) 
\left\langle v_{l}^{*}, q^{L_{0}} v_{m} \right\rangle_{{\cal V}_{\beta}} 
{\cal F}_{C_{2}} (v_{m}^{*} \otimes v_{2}), 
$$
where $\{ v_{l} \}$, $\{ v_{l}^{*} \}$ are dual bases of ${\cal V}_{\beta}$ 
for $\langle \cdot, \cdot \rangle_{{\cal V}_{\beta}}$. 
Then ${\cal F}_{C_{1} \sharp C_{2}}^{\beta} (v_{1} \otimes v_{2})$ is the product of  $q^{\Delta_{\beta}}$ and a formal power series of $q$ whose constant term is 
${\cal F}_{C_{1}} (v_{1} \otimes e_{\beta}) {\cal F}_{C_{2}} (e_{\beta} \otimes v_{2})$. 
For a Riemann surface $C$ with $n+2$ marked points and 
associated local coordinates, 
the gluing ${\cal F}_{C^{\sharp}}^{\beta}$ of the conformal block ${\cal F}_{C}$ 
by $\beta \in {\mathbb S}$ can be defined in a similar way, 
where $C^{\sharp}$ denotes the $n$-pointed Riemann surface obtained by 
gluing the $(n+1), (n+2)$th points on $C$. 
Let $\sigma$ be a pants decomposition of a Riemann surface 
of genus $g$ with $n$ marked points and local coordinates, 
and $\beta$ be an ${\mathbb S}$-valued function on the set $E(\sigma)$ of edges 
associated with $\sigma$. 
Then we define the {\it gluing conformal block} ${\cal F}_{\sigma}^{\beta}$ 
as the gluing of conformal blocks on $3$-pointed Riemann spheres, 
and it is represented as a formal power series of $q_{e}$ $(e \in E(\sigma))$ 
multiplied by $\prod_{e \in E(\sigma)} q_{e}^{\Delta_{\beta(e)}}$, 
where $q_{e}$ denote deformation parameters of 
the degenerate curve associated with $\sigma$. 

Let $C/S$ be a family of stable curves over ${\mathbb C}$ of genus $g$ 
with $n$ marked points $p_{i}$ and local coordinates $z_{i}$ $(1 \leq i \leq n)$. 
Denote by $\sigma_{i}: S \rightarrow C$ the section corresponding to $p_{i}$. 
Then it is shown in \cite[7.4]{BK2} that in the category of algebraic geometry, 
one can let ${\cal T}_{S}$ act on the sheaf of conformal blocks as follows. 
For a vector field $\theta$ on $S$, 
there exists a lift $\tilde{\theta}$ as a vector field on 
$C - \bigcup_{i=1}^{n} \sigma_{i}(S)$ since it is affine over $S$. 
Take the $i$th vertical component (for the local coordinate $z_{i}$) 
$\tilde{\theta}_{i}^{\rm vert}$ of $\tilde{\theta}$ as 
$$
\tilde{\theta} = \tilde{\theta}_{i}^{\rm vert} + \tilde{\theta}_{i}^{\rm holiz}; 
\ \ \tilde{\theta}^{\rm holiz}(z_{i}) = 0. 
$$
Namely, if $\theta = \sum_{k} f_{k}(x) \partial/\partial x_{k}$ and 
$\tilde{\theta} = 
\sum_{k} f_{k}(x) \partial/\partial x_{k} + g(x, z_{i}) \partial/\partial z_{i}$, 
where $x = (x_{k}), z_{i}$ are local coordinates around $\sigma_{i}(S)$, 
then $\tilde{\theta}_{i}^{\rm vert} = g(x, z_{i}) \partial/\partial z_{i}$.  
Define the action of $\tilde{\theta}$ on 
${\cal O}_{S} \otimes \bigotimes_{i=1}^{n} {\cal V}_{\alpha_{i}}$ as 
$$
\tilde{\theta} (f \otimes v) = 
\theta(f) \otimes v + f \otimes \sum_{i=1}^{n} \rho_{\tilde{\theta}_{i}^{\rm vert}} (v) 
\ \ \left( f \in {\cal O}_{S}, \ v \in \bigotimes_{i=1}^{n} {\cal V}_{\alpha_{i}} \right). 
$$
Then by the definition of conformal blocks, 
this action gives the action of ${\cal T}_{S}$ on the sheaf of conformal blocks on $C/S$. 
We denote $\nabla$ the corresponding connection. 
Then it is known that $\nabla$ is projectively flat, 
the residue of $\nabla$ around the singular locus of $C/S$ is given by 
the $L_{0}$-eigenvalue 
and that ${\cal F}_{\sigma}^{\beta}$ gives a (formal) flat section of $\nabla$ 
(cf. \cite[Sections 7.4 and 7.8]{BK2}).  

\subsection{Modular functor} 

We recall results of Teschner \cite{T1,T2,T3} on analytic continuations of 
Liouville conformal blocks on $4$-pointed Riemann spheres. 
We normalize 
$N(\alpha_{1}, \alpha_{2}, \alpha_{3}) = 
{\cal F}_{C} \left( e_{\alpha_{1}} \otimes e_{\alpha_{2}} \otimes e_{\alpha_{3}} \right)$ 
as in \cite[(8.3) and (12.22)]{TV}, 
and $\sigma$, $\sigma'$ be pants decompositions of 
${\mathbb P}^{1}_{\mathbb C} - \{ 0, 1, \infty, x\}$ which are connected by 
a fusing move $x \in (0, 1)$. 
Then it was shown in \cite{T1,T2,T3} that for each $\beta \in {\mathbb S}$, 
the associated conformal block 
$$
{\cal F}_{\sigma}^{\beta} : 
{\cal V}_{\alpha_{1}} \otimes {\cal V}_{\alpha_{2}} \otimes 
{\cal V}_{\alpha_{3}} \otimes {\cal V}_{\alpha_{4}} 
\rightarrow {\mathbb C}
$$
can be analytically continued along $(0, 1)$ to 
a meromorphic form around $x = 1$ which is represented as 
$$
\int_{\mathbb S} d \mu(\beta') \ \Phi_{\beta, \beta'} \ {\cal F}_{\sigma'}^{\beta'} 
$$
for a kernel function $\Phi_{\beta, \beta'}$ and a measure $d \mu(\beta')$ 
which are explicitly described in \cite[5.2]{T2} and \cite[2.1]{T3}. 
Therefore, 
the analytic continuation along $(0, 1)$ gives rise to a canonical isomorphism 
$$
\int_{\mathbb S}^{\oplus} d \beta \ {\mathbb C} {\cal F}_{\sigma}^{\beta} \ \cong \ 
\int_{\mathbb S}^{\oplus} d \beta' \ {\mathbb C} {\cal F}_{\sigma'}^{\beta'} 
$$
between the Hilbert spaces of square-integrable functions on ${\mathbb S}$ 
which are obtained as direct integrals. 

Furthermore, the analytic continuation of ${\cal F}_{\sigma}^{\beta}$ 
along a simple move in ${\cal M}_{1, \vec{1}}^{\rm an}$ was given by 
Hadasz-Jask\'{o}lski-Suchanek \cite{HJS}, 
and that along the half-Dehn twist associated with an edge $e$ is the multiplication by 
$\exp \left( \pi \sqrt{-1} \Delta_{\beta(e)} \right)$ 
on ${\mathbb C} {\cal F}_{\sigma}^{\beta}$. 

Let $C$ be a Riemann surface of genus $g$ with $n$ marked points $p_{i}$ and 
local coordinates $z_{i}$ $(1 \leq i \leq n)$. 
Take a pants decomposition $\sigma$ of $C$, 
and denote by $E(\sigma)$ the set of edges associated with $\sigma$. 
We consider the deformation of the maximally degenerate pointed curve 
obtained by pinching the real curves in $\sigma$, 
and its deformation by the parameters $q_{e}$ $(e \in E(\sigma))$ given in 3.1. 
Let $\beta$ be an ${\mathbb S}$-valued function on $E(\sigma)$. 
Then by the definition of the gluing conformal block ${\cal F}_{\sigma}^{\beta}$, 
for each $v \in \bigotimes_{i=1}^{n} {\cal V}_{\alpha_{i}}$, 
one can see that 
$$
\prod_{e \in E(\sigma)} q_{e}^{-\Delta_{\beta(e)}} \cdot {\cal F}_{\sigma}^{\beta} (v) 
$$
becomes a formal power series of $q_{e}$ $(e \in E(\sigma))$. 
Denote by $C_{\sigma}^{\beta} (v)$ its constant term. 
Let $p_{C}$ denote the point on ${\cal M}_{g, \vec{n}}^{\rm an}$ 
corresponding to $(C; p_{i}, z_{i})$, 
and $p_{\infty}$ denote the tangential base point over ${\mathbb Z}$ at infinity 
obtained by taking $q_{e}$ $(e \in E(\sigma))$ as sufficiently small positive numbers. 
Take a smooth path $\pi$ in $\overline{\cal M}_{g, \vec{n}}^{\rm an}$ 
from $p_{C}$ to $p_{\infty}$, 
and consider the parallel transport ${\rm Trans}_{p_{C}}^{p}$ by $\nabla$ along $\pi$ 
from $p_{C}$ to $p \in \pi$. 
Then one can define the conformal block ${\cal F}_{C}^{\beta}$ 
associated with $(C; p_{i}, z_{i}, \pi)$ as a linear map 
$\bigotimes_{i=1}^{n} {\cal V}_{\alpha_{i}} \rightarrow {\mathbb C}$ satisfying 
$$
\lim_{t \downarrow 0} \prod_{e \in E(\sigma)} t^{-\Delta_{\beta(e)}} \cdot 
{\rm Trans}_{p_{C}}^{p(t)} \left( {\cal F}_{C}^{\beta} (v) \right) 
= C_{\sigma}^{\beta} (v), 
$$
where $t \mapsto p(t) \in \pi$ is a smooth map for sufficiently small $t > 0$ 
such that for each $e \in E(\sigma)$, 
the local coordinate $q_{e}(t)$ of $p(t)$ belongs to 
${\mathbb R}_{+} = \{ r \in {\mathbb R} \ | \ r > 0 \}$
and $\lim_{t \downarrow 0} q_{e}(t)/t = 1$. 
Then we have the following theorem which shows the nontrivial assumption in \cite[3.2.2]{T5}  for Liouville conformal blocks.  
\vspace{2ex}

\noindent
{\bf Theorem 5.1.} 
\begin{it}
The gluing conformal block ${\cal F}_{\sigma}^{\beta}(v)$ is convergent and 
analytically continued as a multi-valued holomorphic function 
on ${\cal M}_{g, \vec{n}}^{\rm an}$ given by ${\cal F}_{C}^{\beta} (v)$. 
\end{it} 
\vspace{2ex}

We call ${\cal F}_{C}^{\beta}$ {\it normalized} if ${\cal F}_{\sigma}^{\beta}$ is constructed 
from normalized conformal blocks on $3$-pointed Riemann spheres. 
Then we can define the space 
${\rm CB}^{\rm temp} \left( \otimes_{i=1}^{n} {\cal V}_{\alpha_{i}}, C \right)$ 
of {\it tempered} conformal blocks associated with $(C; p_{i}, z_{i})$ given by 
the direct integral 
$$
\int_{{\mathbb S}^{3g-3+n}}^{\oplus} \prod_{e \in E(\sigma)} d \beta(e) \ 
\bigotimes {\mathbb C} {\cal F}_{C}^{\beta}, 
$$
where ${\cal F}_{C}^{\beta}$ are normalized. 
This space is isomorphic to the Hilbert space of square-integrable functions on 
$$
\left\{ (\beta(e))_{e \in E(\sigma)} \ | \ \beta(e) \in {\mathbb S} \right\} 
\cong ({\mathbb R}_{+})^{3g-3+n}. 
$$

\noindent
{\bf Theorem 5.2.} 
\begin{it}
\begin{itemize}

\item[\rm (1)]
The Hilbert space 
${\rm CB}^{\rm temp} \left( \otimes_{i=1}^{n} {\cal V}_{\alpha_{i}}, C \right)$ 
is independent of $\pi$ and $p_{\infty}$. 

\item[\rm (2)] 
The Hilbert space 
${\rm CB}^{\rm temp} \left( \otimes_{i=1}^{n} {\cal V}_{\alpha_{i}}, C \right)$ 
satisfies the factorization rule in the following sense. 
For Riemann surfaces $C_{k}$ $(k = 1, 2)$ with $n_{k} + 1$ marked points 
and local coordinates, 
$$
{\rm CB}^{\rm temp} \left( \left( \otimes_{i=1}^{n_{1}} {\cal V}_{\alpha_{i}} \right) 
\otimes \left( \otimes_{j=1}^{n_{2}} {\cal V}_{\alpha_{j}} \right), 
C_{1} \sharp C_{2} \right)
$$
is canonically isomorphic to the direct integral     
$$
\int_{\mathbb S}^{\oplus} d \beta \ 
{\rm CB}^{\rm temp} \left( \left( \otimes_{i=1}^{n_{1}} {\cal V}_{\alpha_{i}} \right) 
\otimes {\cal V}_{\beta}, C_{1} \right) \otimes 
{\rm CB}^{\rm temp} \left( {\cal V}_{\beta} \otimes 
\left( \otimes_{j=1}^{n_{2}} {\cal V}_{\alpha_{j}} \right), C_{2} \right). 
$$
Similarly, for a Riemann surface $C$ with $n + 2$ marked points and local coordinates, 
one has a canonical isomorphism 
$$
{\rm CB}^{\rm temp} \left( \otimes_{i=1}^{n} {\cal V}_{\alpha_{i}}, C^{\sharp} \right) 
\cong 
\int_{\mathbb S}^{\oplus} d \beta \ 
{\rm CB}^{\rm temp} \left( \left( \otimes_{i=1}^{n} {\cal V}_{\alpha_{i}} \right) 
\otimes {\cal V}_{\beta} \otimes {\cal V}_{\beta}, C \right). 
$$

\item[\rm (3)]
By the connection $\nabla$, 
${\rm CB}^{\rm temp} \left( \otimes_{i=1}^{n} {\cal V}_{\alpha_{i}}, C \right)$ 
has a projective action of the Teichm\"{u}ller groupoid for ${\cal M}_{g, \vec{n}}$ 
such that the action of fusing moves and simple moves is given by the action 
in the case when $(g, n) = (0, 4)$ and $(1, 1)$ respectively. 

\item[\rm (4)] 
The action on ${\rm CB}^{\rm temp}$ by basic moves satisfies the 
Moore-Seiberg equations (cf. \cite{MS}, \cite[6.6.1]{TV}) up to nonzero constants. 

\end{itemize}
\end{it}

\noindent 
{\it Proof.} 
First, we prove (1). 
Since $\nabla$ is projectively flat, 
${\rm CB}^{\rm temp} = 
{\rm CB}^{\rm temp} \left( \otimes_{i=1}^{n} {\cal V}_{\alpha_{i}}, C \right)$ 
is independent of the homotopy class of $\pi$. 
Then by Theorem 3.1, to prove (1), 
it is enough to show that ${\rm CB}^{\rm temp}$ is independent of 
moving $p_{\infty}$ by fusing moves and simple moves. 
Let $\sigma$ and $\sigma'$ be pants decompositions of Riemann surfaces of genus $g$ 
with $n$ marked points such that $\sigma$, $\sigma'$ are connected by a fusing move $\varphi$. 
Then a gluing conformal block ${\cal F}_{\sigma}^{\beta}$ is represented as 
the gluing ${\cal F}_{C_{1} \sharp C_{2}}^{\beta}$ of ${\cal F}_{C_{1}}^{\beta_{1}}$ and 
${\cal F}_{C_{2}}^{\beta_{2}}$, 
where $C_{1}$ denotes a $4$-pointed Riemann sphere associated with $\varphi$. 
By the above result of Teschner \cite{T1, T2, T3}, 
there exists a form ${\cal F}'_{C_{1}}$ which is the parallel transport of 
${\cal F}_{C_{1}}^{\beta_{1}}$ along the fusing move in ${\cal M}_{0, \vec{4}}^{\rm an}$ 
associated with $\varphi$. 
As is shown in \cite[3.1]{I2} and \cite[1.2]{I3}, 
there exist local coordinates $x, u_{1},..., u_{3g+n-4}$ of the point at infinity 
on $\overline{\cal M}_{g,n}$ corresponding to $\sigma$ such that 
$1-x, u_{1},..., u_{3g+n-4}$ give local coordinate of the point at infinity 
corresponding to $\sigma'$ and that $0 < x < 1$ gives the fusing move $\varphi$. 
Then the parallel transport of ${\cal F}_{\sigma}^{\beta}$ along $\varphi$ becomes 
the gluing of ${\cal F}'_{C_{1}}$ and ${\cal F}_{C_{2}}^{\beta_{2}}$ 
by the deformation parameters $u_{i}$ given in Theorem 2.2. 
Therefore, ${\rm CB}^{\rm temp}$ is independent of moving $p_{\infty}$ by fusing moves. 
By the result of Hadasz-Jask\'{o}lski-Suchanek \cite{HJS}, 
the space of tempered conformal blocks for $1$-pointed curves of genus $1$ is 
stable under a simple move. 
Therefore, in a similar way as above, 
one can show that ${\rm CB}^{\rm temp}$ is independent of moving $p_{\infty}$ 
by simple moves.  

Second, we prove (2) in the former case (and the latter case can be shown in a similar way). 
Take pants decompositions of the Riemann surfaces $C_{i}$ $(i = 1, 2)$ 
which give a pants decomposition of $C_{1} \sharp C_{2}$, 
and denote by $p_{\infty}$ the associated tangential point $p_{\infty}$ at infinity. 
Then one can obtain the required isomorphism from the description of 
the space of tempered conformal blocks  associated with $C_{1} \sharp C_{2}$ by $p_{\infty}$. 

The assertion (3) follows from the construction of 
the space of tempered conformal blocks and the proof of (1),  (2). 

Finally, the assertion (4) follows that the action of basic moves gives rise to 
a projective representation of the mapping class group. 
\ $\square$ 
\vspace{2ex}

By this theorem, one can see that ${\rm CB}^{\rm temp}$ gives 
a generalized modular functor with label set ${\mathbb S}$ and identity involution 
$\hat{\alpha} = \alpha$ $(\alpha \in {\mathbb S})$. 
\vspace{2ex}

\noindent
{\bf Theorem 5.3}  
\begin{it} 
Let the notation be as above. 
Then $\prod_{e \in E(\sigma)} |q_{e}|^{-\Delta_{\beta(e)}} {\cal F}_{\sigma}^{\beta}(v)$ 
gives the wave function with initial value $C_{\sigma}^{\beta}(v)$ 
of the Hilbert bundle associated with ${\rm CB}^{\rm temp}$ and 
conformal dimensions $r_{\alpha} = \Delta_{\alpha}$ $(\alpha \in {\mathbb S})$. 
\end{it} 
\vspace{2ex}

\noindent
{\it Proof.} 
This assertion follows from Theorem 4.3 and the above property of 
${\cal F}_{\sigma}^{\beta}(v)$. 
\ $\square$

\section{Quantum Teichm\"{u}ller theory} 

Verlinde \cite{V} posed a conjecture that the quantum Teichm\"{u}ller theory 
(cf. \cite{CF, K1, K2, K3}) gives a generalized modular functor 
which is equivalent to the Liouville modular functor reviewed in Section 5. 
Teschner \cite{T3, T4, T6} and Teschner-Vartanov \cite{TV} showed this conjecture 
as follows. 
\begin{itemize}

\item 
Let $\Sigma$ be a hyperbolic surface of genus $g$ and $n$ geodesic boundaries, 
and denote by $\Gamma$ the dual graph of a pants decomposition of $\Sigma$. 
Then there exist quantum length operators $\mbox{\boldmath $l$}_{i}$ 
$(1 \leq i \leq 3g-3+n)$ corresponding to the geodesic length functions $l_{i}$, 
and a Hilbert space ${\cal H}^{\rm T}(\Sigma)$ with basis 
$$
{\cal B}_{\Gamma} = \left\{ \left| \Gamma_{\Lambda, L} \right\rangle \ | \ 
L = (l_{1},..., l_{3g-3+n}) \in ({\mathbb R}_{+})^{3g-3+n} \right\}
$$ 
of generalized eigenfunctions of 
$\left\{ \mbox{\boldmath $l$}_{1},..., \mbox{\boldmath $l$}_{3g-3+n} \right\}$. 

\item 
Each basic move $\Gamma \mapsto \Gamma'$ between pants decompositions gives 
an integral representations of ${\cal B}_{\Gamma}$ by  ${\cal B}_{\Gamma'}$ 
which is equivalent to the representation in the Liouville theory given in 5.2 by putting 
$$
\alpha_{i} = \frac{Q}{2} + \sqrt{-1} \frac{l_{i}}{4 \pi b}; \ 
Q = b + \frac{1}{b}. 
$$

\item 
The operators given by basic moves give a projective representation 
on the Teichm\"{u}ller groupoid with central extension $\gamma$ which is related to 
the Liouville central charge 
$c = 1 + 6 Q^{2}$ by $\gamma = \exp \left( \frac{\pi \sqrt{-1}}{2 c} \right)$. 

\end{itemize} 

Then by Theorems 4.2 and 5.3, 
one can show Conjecture 5.1 (ii) of \cite{T3} that the Liouville conformal blocks 
represent the generalized eigenfunctions of the length operators 
in the quantum Teichm\"{u}ller theory under the above relation of parameters.






\end{document}